\documentclass[aps,prb,twocolumn,superscriptaddress,letter]{revtex4}
\usepackage{graphicx,epsf}

\begin{document}

\title{Polarization-selective excitation of N-V centers in diamond}

\author{T. P. Mayer Alegre}
\affiliation{Laborat\'{o}rio Nacional de Luz S\'{\i}ncrotron, Caixa
Postal 6192 - CEP 13083-970, Campinas, SP, Brazil}
\affiliation{Instituto de F\'{\i}sica Gleb Wataghin, Universidade
Estadual de Campinas, Campinas, SP, Brazil}

\author{C. Santori}
\affiliation{Hewlett-Packard Laboratories, 1501 Page Mill Rd., Palo
Alto, California 94304, USA}

\author{G. Medeiros-Ribeiro}
\affiliation{Laborat\'{o}rio Nacional de Luz S\'{\i}ncrotron, Caixa
Postal 6192 - CEP 13083-970, Campinas, SP, Brazil}
\affiliation{Hewlett-Packard Laboratories, 1501 Page Mill Rd., Palo
Alto, California 94304, USA}

\author{R. G. Beausoleil}
\affiliation{Hewlett-Packard Laboratories, 1501 Page Mill Rd., Palo
Alto, California 94304, USA}

\date{\today}
\begin{abstract}
The nitrogen-vacancy (N-V) center in diamond is promising as an
electron spin qubit due to its long-lived coherence and optical
addressability.  The ground state is a spin triplet
with two levels ($m_s = \pm 1$) degenerate at
zero magnetic field.  Polarization-selective microwave excitation
is an attractive method to address the spin transitions
independently, since this allows operation down to zero magnetic
field.  Using a resonator designed to produce circularly polarized
microwaves, we have investigated the polarization selection
rules of the N-V center.  We first apply this technique to N-V
ensembles in [100] and [111]-oriented samples.  Next, we demonstrate
an imaging technique, based on optical polarization dependence, that
allows rapid identification of the orientations of many single N-V
centers.  Finally, we test the microwave polarization selection rules
of individual N-V centers of known orientation.
\end{abstract}

\date{\today}

\maketitle
\section{Introduction}

Nitrogen-vacancy (N-V) centers in diamond are a promising system for
implementing and testing quantum information processing protocols in
solids.  Spin states of single centers can be initialized and
detected optically~\cite{gruber,jelezko1,jelezko2,jelezko3,hanson1},
and this capability has been extended to controlled coupling between
N-V centers and other spins
($^{13}C$,$N$)~\cite{wrachtrup,jelezko4,gaebel,hanson2,childress}. A
special feature of this system is the spin-triplet structure of the
ground states, with individual transitions that can in theory be
manipulated independently using microwave polarization selection
rules. The ground state of the negatively charged N-V center is
known to have a ($^{3}A$) spin triplet structure~\cite{van-Oort1}
which, neglecting hyperfine interactions, is described by the
following electron spin Hamiltonian~\cite{loubser}:
\begin{equation}\label{eq_electron}
H_{e}=D (S_{z}^{2}+\frac{1}{3}S^2)+g\beta\mathbf{S}\cdot\mathbf{B}
\,,
\end{equation}
where $\mathbf{B}$ is the magnetic field,
$g\beta=27.98\,\mathrm{GHz/T}$ and $D=2.88\,\mathrm{GHz}$ is the
zero-field splitting between the lowest-energy $m_s=0$ sublevel, and
the $m_s=\pm1$ sublevels which are degenerate at zero magnetic
field.  In typical optically-detected magnetic resonance (ODMR)
experiments, the microwave transitions are driven by linearly
polarized fields, and individual transitions are selected by
applying a magnetic field to lift the $m_s=\pm1$ degeneracy.  To
avoid spin mixing, the magnetic field is usually applied along the
quantization axis, which can be along any of the four
$\langle111\rangle$ crystal axes.

An alternative method to drive the spin transitions selectively is
to take advantage of polarization selection rules, since the
transitions from $m_s=0$ to $m_s=\pm1$ occur with
$\sigma^{\pm}$ circularly polarized microwave excitation.  Using
polarization, rather than frequency, selective excitation at zero
magnetic field becomes possible, and
at weak magnetic fields where transitions still overlap
(\textit{e.g.} $\rm{B}\leq5\rm{G}$) the selectivity can be
improved. At small magnetic fields and under weak strain or electric
field, it is expected that simultaneous cycling and
non-spin-conserving optical transitions can be realized, useful
for optical readout and manipulation, respectively~\cite{manson,tamarat}.
Therefore, high selectivity and weak magnetic fields are highly
desired for improving the prospects of N-V-based quantum information
processing.

Here we demonstrate selective excitation of ensembles and individual
N-V centers using circularly polarized microwaves for the spin
transitions, and the laser polarization with respect to the N-V axis
for the optical transitions.

\begin{figure}[tp]
\centerline{\epsffile{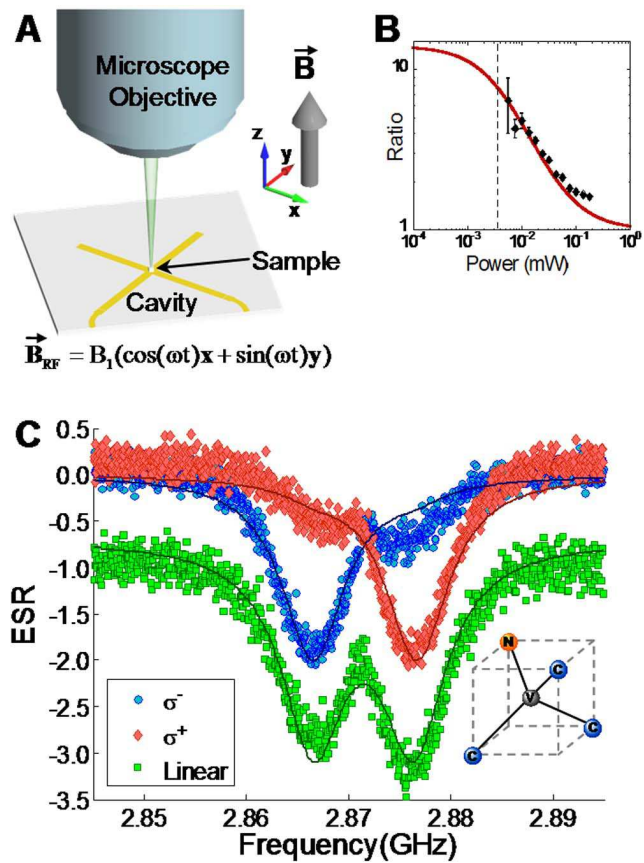}} \caption{\textbf{ODMR
experiment on $(100)$ sample} \textbf{a.} Schematic drawing showing
the experimental positioning of the sample relative to the cavity
and the magnetic field direction, which is also the rotation axis
for the ac magnetic field. \textbf{b.} ODMR ratio
($\sigma^+/\sigma^-$ amplitude) vs. microwave power intensity
(points), fitted empirical equation as described in text (solid
line), and saturation power (dashed line). \textbf{c.} Ensemble ODMR
spectra for three microwave polarizations: $\sigma^-$ (blue
circles), $\sigma^+$ (red diamonds) and linear (green squares). The
solid lines are fits of gaussian peaks with adjustable amplitudes
and $5\,\mathrm{MHz}$ width. The inset is a schematic drawing
showing the 4 possible N-V orientations.} \label{sample100}
\end{figure}

\section{Experimental Setup}

To test the selection rules experimentally, we performed
optically-detected magnetic resonance (ODMR) experiments using a
microstrip resonator designed to produce circularly polarized
microwaves
.  The cavity is formed by two crossed microstrip resonators with
orthogonal modes which generate microwaves with any polarization
(from linear to circular) when the two inputs are driven with a
particular phase difference ($0^{\circ}$ to $90^{\circ}$).  Our
cavity has a full-width half maximum (FWHM) resonance of
$30~\mathrm{MHz}$ with a quality factor of $Q=100$.  To avoid
distortions from the resonance shape of the cavity, the measurements
were kept within the FWHM response of the cavity. After positioning
the sample above the cavity center, the resonance frequency of the
cavity was trimmed by
placing a low-loss dielectric above it. Three different microwave
polarization configurations were used: linearly polarized, clockwise
($\sigma^{-}$) and counterclockwise ($\sigma^{+}$) circular
polarizations with respect to the static magnetic field.  The static
magnetic field was produced by a permanent magnet placed in a linear
translator close to the sample. The magnetic field direction was set
along the optical detection axis (perpendicular to sample surface).
Apart from the microwave polarization control, the techniques used
here are similar to those used in previous room-temperature ODMR
experiments~\cite{gruber,van-Oort1}. Fig.~\ref{sample100}(a) shows a
schematic for our experimental setup, showing the sample position
above the cavity, as well as the cavity design.  We used a confocal
setup in which an excitation laser ($532\,{\rm nm}$) was focused to
a sub-micron spot inside the sample for non-resonant excitation of
the N-V centers. This setup also allowed for the control of the
laser polarization incident on the sample.  Laser excitation
produces a spin polarization, preferentially
populating the $m_s=0$ ground state. When the applied microwave
field with the appropriate polarization is resonant with one of the
spin transitions, the spin polarization decreases, producing a
measurable decrease in the photoluminescence intensity.

\section{Results and Discussion}
\subsection{$[100]$ Oriented Sample: Ensemble Measurements}

The first set of measurements (Fig.~\ref{sample100}) was performed
on a $[100]$ oriented sample with a relatively high $(\sim 10^2 \,
\mu\mathrm{m}^{-2})$ concentration of N-V centers near the surface.
For this crystal orientation, none of the four possible N-V
orientations (Fig.~\ref{sample100}(c), inset) have their
quantization axes parallel to the surface normal, and hence a
circular polarization in the laboratory frame of reference has an
elliptical projection onto a plane perpendicular to the N-V axis.
However the ODMR spectrum is the simplest for this crystal
orientation because for each N-V orientation the absolute angle
between the quantization axis and the surface normal is the same.
The normalized transition probabilities between $m_s=0$ to $m_s\pm1$
are calculated theoretically to be $(2+\sqrt{3})/12$ and
$(2-\sqrt{3})/12$, respectively.

For a quantitative verification of the selection rules we would
ideally perform measurements in a regime where the microwave
power dependence of the ODMR spectrum is linear.  The ODMR signal
dependence on the microwave power excitation can be
phenomenologically described by $A=A_{max}/(1+P_{sat}/P)$, where
$P_{sat}$ is the -3dB saturation level.  Figure \ref{sample100}(b)
shows a measurement of the resonance amplitude ratio between the two
transitions ($m_s=0 \rightarrow m_s=+1$ and
$m_s=0\rightarrow~m_s=-1$) as a function of applied microwave power.
The solid line is a fit using the ratio between the two transitions,
$R = (P+P_{\rm sat}^{\rm \sigma^{\mp}})/(P+P_{\rm sat}^{\rm
\sigma^{\pm}})$ (for the $\sigma^{\pm}$ microwave polarization),
where in the low power limit the ratio is $P_{\rm sat}^{\rm
\sigma^{\mp}}/P_{\rm sat}^{\rm \sigma^{\pm}} =
(2+\sqrt{3})/(2-\sqrt{3})$. With these assumptions the saturation
power for $\sigma^+$ is $\sim-49\mathrm{dBm}$.  Our measurements
were done at $-40\mathrm{dBm}$, since the time required to perform
the experiment below the saturation limit at the same signal to
noise ratio would be prohibitively long.

Figure \ref{sample100}(c) shows ESR spectra measured for three
microwave polarizations, showing two peaks corresponding to the
$m_s=\pm1$ transitions.  The solid lines are the calculated spectra
solving Eq.~\ref{eq_electron} over the 4 possible N-V orientations,
considering an inhomogeneous broadening of $7.5~\mathrm{MHz}$
(obtained from the measured data) for each transition.  The model
does not include the saturation behavior.  Circularly
polarized microwave excitation produced a constrast ratio greater
than $75\%$ between the two ODMR peaks. In the linearly polarized
spectrum (shifted for clarity), the slight asymmetry
is due to the cavity response.

\begin{figure}[!h]
\centerline{\epsffile{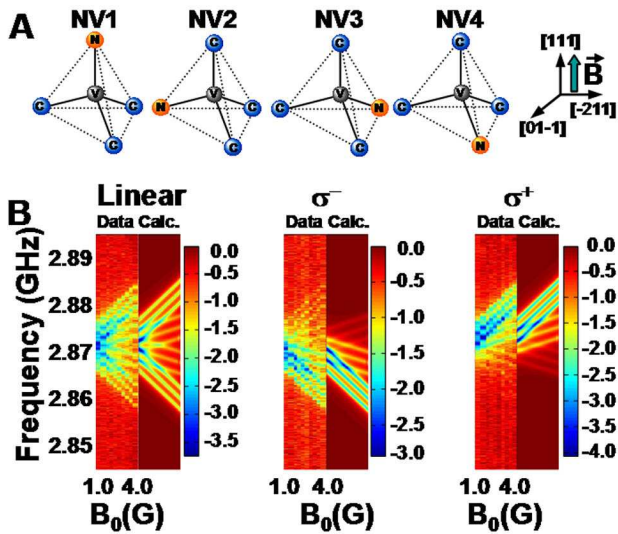}}
\caption{\textbf{Ensemble ODMR spectra for $[111]$ oriented sample
showing hyperfine interaction.} \textbf{a.} Schematic drawing
showing the relation between the 4 possible N-V orientations and the
magnetic field direction. \textbf{b.} Color plot showing ODMR
spectra vs. magnetic field (horizontal axis, from 1 to 4 gauss) and
microwave frequency (vertical axis, $\rm{50MHz}$ range) for linear,
$\sigma^-$ and $\sigma^+$ microwave polarizations, comparing data
(left) and simulation (right) as described in the text. The
simulation uses an inhomogeneous broadening of $1.2$MHz to fit the
data. The color scale indicates the peak depth, in percent, relative
to the off-resonant case. The average number of counts, for the
three measurements, is $4\times10^{5}$.} \label{hyperfine}
\end{figure}

\begin{figure*}[tp]
\centerline{\epsffile{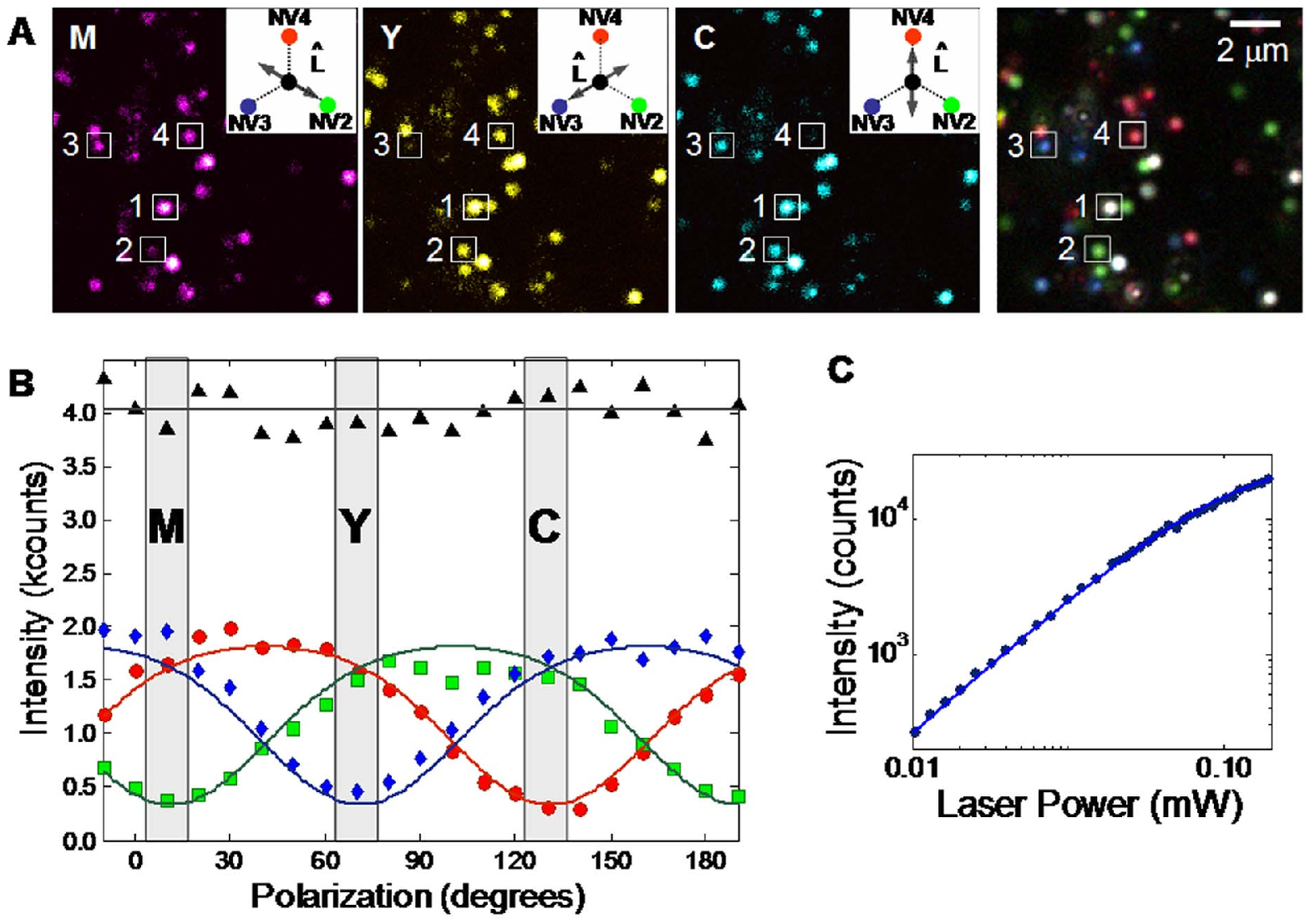}} \caption{
\textbf{Determination of N-V orientation through optical
polarization dependence.} \textbf{a.} Scanning confocal
photoluminescence images showing single N-V centers, with laser
polarization encoded into the three color channels MYC (magenta,
yellow, and cyan). The first three images were measured for
particular polarizations in which orientations designated as NV2,
NV3, and NV4 have photoluminescence minima, with the electric field
nearly parallel to the N-V axis. A representative N-V center of each
orientation is also labeled. The fourth image is a sum over many
measurement polarization angles, in which orientations NV1, NV2, NV3
and NV4 appear as white, green, blue and red, respectively.
\textbf{b.} Luminescence intensity vs. excitation polarization for
selected single N-V centers with orientations NV1 (black triangles,
oriented normal to sample surface), NV2 (green squares), NV3 (blue
diamonds), and NV4 (red circles). These measurements were performed
at weak excitation power to avoid saturation of the optical
transition, which distorts the polarization dependence. \textbf{c.}
Excitation power dependence for a single N-V center with NV1
orientation, fitted to a simple model (solid line). }
\label{laser_pol}
\end{figure*}

\subsection{$[111]$ Oriented Sample: Ensemble Measurements}
Next, we describe measurements performed on a $[111]$-oriented
sample with a much lower N-V concentration, below $1 \,
\mu\mathrm{m}^{-3}$. For this crystal orientation, from four
possible N-V orientations (see Fig.~\ref{hyperfine}(a)), one of them
($[111]$, labeled ``NV1'') is normal to the sample surface and is
expected theoretically to exhibit perfect selection rules. For the
other orientations the projected microwave field is elliptical in
the N-V reference frame, and the circular selection rules are not
fully observed.  Thus, this measurement can reveal the orientation
of a particular center. For this experimental configuration, the
Zeeman splitting depends on the N-V orientation, decreasing by a
factor of approximately $3$ for N-V orientations other than $[111]$.

Due to the high purity of this sample, hyperfine structure can be
resolved in the ODMR spectrum.  For a substitutional $^{14}N$ with
nuclear spin $I=1$, the Hamiltonian terms describing the nuclear
Zeeman, hyperfine and quadrupole interactions are~\cite{forrest}:
\begin{eqnarray}\label{eq_hyperfine}\nonumber
H_{T}= P(I_{z}^{2}-\frac{1}{3}I^{2}) -
(g_{n}\beta_{n})\mathbf{B}\cdot\mathbf{I}
+...\\
+ A_{\parallel} S_{z} I_z +\frac{1}{2}A_{\perp}(S^{-}I^{+} + S^{+}I^{-})
\,,
\end{eqnarray}
where $\beta_{n}$ is the nuclear magneton, $A_{\parallel}$ and
$A_{\perp}$ are the hyperfine constants and $P$ is the quadrupole
splitting.  Previous work~\cite{forrest,xing-fei1,xing-fei2}
has established these constants to be:
$A_{\parallel}=2.3\,\mathrm{MHz}$, $A_{\perp}=2.1\,\mathrm{MHz}$,
$g_{n}\beta_{n}=3.07\,\mathrm{MHz}/\mathrm{T}$ and
$P=-5.04\,\mathrm{MHz}$. For small static magnetic fields, the
hyperfine interaction splits each of the two ODMR peaks into a
triplet with splittings of approximately
$A_{\parallel} = 2.3\,{\rm MHz}$. In this
limit, each component of a triplet follows approximately the
original circular polarization selection rules.

The ensemble ODMR spectra in Fig.~\ref{hyperfine}(b), which show the
hyperfine structure, were obtained by removing the pinhole from the
collection optics to reduce spatial selectivity, and exciting with
small optical and microwave powers. Measured and calculated spectra
using Eq.~\ref{eq_hyperfine} are shown. The calculation includes all
possible transitions, weighted by their predicted strengths, using a
lorentz function ($1.2~\mathrm{MHz}$ linewidth) to account for
inhomogeneous broadening.  For each panel in \ref{hyperfine}(b), the
color scale represents the ESR intensity as a function of the static
magnetic field (x-axis) and the microwave frequency (y-axis) for the
three different microwave polarizations.  Two different slopes are
observed for the Zeeman splitting, corresponding to N-V centers
oriented along $[111]$ (steeper slope) and the other three
orientations. These results exhibit good agreement between theory
and experiment. 
For the N-V centers oriented along $[111]$, the measured contrast
ratio for $\sigma^+$ vs. $\sigma^-$ excitation is approximately 9.

\subsection{Optical determination of defect orientation}

For work on single N-V centers, it is helpful to have
a simple method to determine the orientation of the
N-V axis.  Here, we demonstrate a method based on optical
polarization dependence.  Since the dipole moments of the optical
transitions are perpendicular to the N-V axis, the optical
polarization anisotropy depends on orientation~\cite{epstein}.
For a $[111]$ sample,
all four orientations can be distinguished optically, allowing
rapid identification of many N-V centers through polarized scanning
confocal imaging.  For effecient representation of the data,
we combine many
images obtained with different excitation polarizations into a
single image by encoding polarization into color according to the
following filter function:
\begin{equation}\label{eq_filter}
f_n(\phi) = a + \sin 2(\phi + \phi_0 + n\pi/3) \, ,
\end{equation}
where $f_n$ is the multiplication factor for color channel $n=0, 1,
2$, $\phi$ is the polarization angle, $a$ is an offset chosen as
small as possible to maximize color saturation without producing
negative pixel values, and $\phi_0$ is an angle offset which
controls hue. This procedure is illustrated in
Fig.~\ref{laser_pol}(a), where the first three panels show images
obtained at the particular polarizations for which N-V orientations
NV2, NV3, and NV4 have photoluminescence minima.  The colors of
these individual images correspond to the encoding in
Eq.~\ref{eq_filter}.  The insets show the relative orientation of
the laser polarization ($\hat{L}$) with respect to these three N-V
axes.  The minima occur when the projection of the N-V axis onto the
sample surface is parallel to the electric field of the optical
excitation.  The last panel shows the combined image in which the
four orientations can be clearly distinguished by color. Based on
photon correlation measurements, we can unambiguously associate the
individual spots with single N-V centers.

The optical polarization dependence
of four representative N-V centers is shown in
Fig~\ref{laser_pol}(b), fitted to a simple
model.  This model includes the effect of N-V orientation on both
the excitation rate and the photoluminescence collection efficiency.
For an N-V center oriented normal to the sample surface (NV1), the
two dipole moments (perpendicular to the N-V axis) can be excited
equally by the laser.  Experimentally, for this orientation we
observe that the photoluminescence is approximately unpolarized,
independent of the excitation polarization (not shown). Therefore in
the model we assume that the two dipole moments of the N-V center
radiate with equal intensity, independent of which dipole moment is
excited by the laser, a situation that would occur if relaxation
between the excited states is fast at room temperature.  When both
polarizations are detected as in Fig~\ref{laser_pol}(b), the other
orientations (NV2, NV3, NV4) produce a weaker photoluminescence
signal even at maximum excitation efficiency since the collection
efficiency is reduced for this geometry.  The model also includes
saturation of the optical transitions (measured in
Fig~\ref{laser_pol}(c)) and asymmetry in the collection efficiency
in our setup for light polarized parallel or perpendicular to the
laser polarization.  Taking all of these factors into account, we
obtain good agreement between theory and experiment.

\begin{figure}[!h]
\centerline{\epsffile{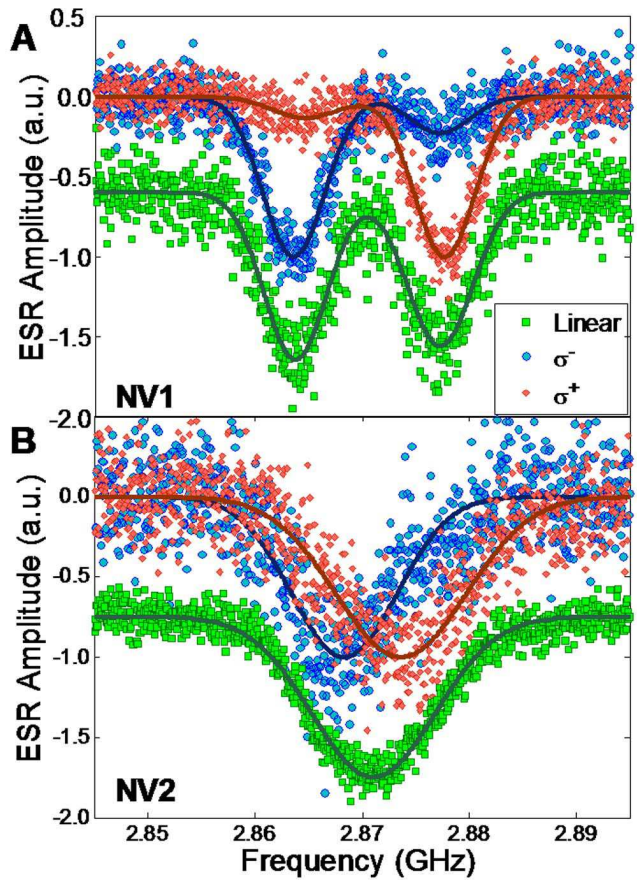}} \caption{\textbf{Single
N-V ODMR measurements.} \textbf{a.} Signal for a single N-V center
oriented along $[111]$ (NV1) for $\sigma^-$ (blue circles),
$\sigma^+$ (red diamonds), and linear (green squares) microwave
polarizations. The solid lines are fits using a sum of two gaussians
with adjustable amplitude. \textbf{b.} Similar measurements
performed on a single center with a different orientation (NV2),
fitted using a single gaussian. The Zeeman splitting is reduced in
this case because of the smaller projection ($\cos \theta = 1/3$) of
the magnetic field onto the N-V axis.} \label{single}
\end{figure}

\subsection{$[111]$ Oriented Sample: Single NV Measurements}
Finally, using the method above to locate single N-V centers of
known orientation, we proceed to test their microwave polarization
selection rules.  The
ODMR spectra for single N-V centers with NV1 and NV2 orientations
are shown in Fig.~\ref{single}, with data (filled symbols) and fits
(solid lines).  In Fig.~\ref{single}(a) the fit used a superposition
of two gaussian curves corresponding to the two microwave
transitions, while in Fig.~\ref{single}(b) the fit used a single
gaussian since, for the smaller Zeeman splitting, the transitions
are unresolved.  For NV1 we expect perfect selectivity,
characterized by a forbidden transition from $m=0$ to $m=-1$ under
$\sigma^+$ excitation.  Experimentally, we observe a residual peak
in this case with approximately 15\% relative intensity which is
most likely due to imperfect generation of circularly polarized
microwave fields above the cavity
, or even the parasitic effect of the trimming dielectric, sample
shape and microscope objective.  Simulations predict that if the
measurement location is just a few millimeters away from the cavity
center, some ellipticity can occur
.   For the NV2
orientation, although the Zeeman splitting is not resolved, exciting
with circular microwave polarization causes a narrowing of the
transition linewidth compared with linearly polarized
excitation, and a shift of the peak center is also observed, showing
that the circular polarization favors one transition.  Theoretically,
we expect the Zeeman splitting for the NV2 orientation to be smaller
by a factor of 3. We also expect reduced polarization selectivity
due to the large angle between the N-V axis and the microwave
polarization axis, with an additional contribution from
the cross-polarized transition of $1/4$ amplitude.

\section{Conclusions}
In summary, we have experimentally verified spin polarization
selection rules for N-V centers in diamond with various crystal and
N-V orientations using circularly polarized microwave excitation.
The results, including hyperfine structure, are consistent with
theory.  We have also demonstrated an efficient optical method for
determining orientations of single N-V centers, and demonstrated
microwave polarization selection rules in single N-V centers of
known orientation.  By optimizing the relative amplitudes and phases
of the signals used to drive the two inputs of the microwave cavity,
a much higher selectivity is possible. Even for the case of the
$[100]$ oriented sample, for a single N-V center, elliptically
polarized excitation could in principle give perfect selectively.
Therefore we expect that microwave polarization selection rules will
prove quite useful for spin manipulation, initialization and
tomography of single N-V centers when improved selectively or
operation at zero static magnetic field are needed.

\section*{ACKNOWLEDGMENTS}

TPMA and GMR acknowledge the financial support of CNPq, FAPESP, and
HP Brazil, and the technical assistance of M. H. Piazzetta and A. L.
Gobbi from the micro-fabrication facility at LNLS. We also
acknowledge Nobuhiko Kobayashi for XRD measurements.  This work was
partially supported by DARPA and the Air Force Office of Scientific
Research through AFOSR contract no.\ FA9550-05-C-0017.

\end{document}